\documentclass[runningheads]{llncs}
\usepackage[T1]{fontenc}
\usepackage{graphicx}
\usepackage{subcaption}
\usepackage{hyperref}
\usepackage{url}
\usepackage{csquotes}
\usepackage{enumitem}
\usepackage{calc}

\usepackage{amsmath}

\usepackage{color}
\usepackage{xcolor}

\usepackage{orcidlink}

\newcommand{\mycomment}[1]{}

\begin{document}
\title{RunSoC 2.0: Scheduling and Allocating Automotive Software Tasks to Hardware Partitions in Heterogeneous MPSoCs\thanks{Preprint submitted to the CASA@ECSA2026 workshop track}}   
\titlerunning{Towards Mixed-Criticality Software Architectures}

%
%
\author{Daniel Krüger\inst{1,2}\orcidlink{0009-0009-8911-9668}\and
Lucas Mauser\inst{1,2}\orcidlink{0000-0002-9235-5145} \and
Stefan Wagner\inst{2}\orcidlink{0000-0002-5256-8429}
}

\institute{Daimler Truck AG, Leinfelden-Echterdingen, Germany \and
Technical University of Munich, Heilbronn, Germany}

%
\authorrunning{D. Krüger et al.}

\maketitle
\begin{abstract}
Centralized automotive architectures increasingly consolidate compute-intensive workloads onto heterogeneous Multi-Processor System-on-Chip (MPSoC), creating strict execution, memory, and communication constraints. This paper presents RunSoC 2.0, a customizable framework for early-stage design-space exploration of task scheduling and allocation on heterogeneous MPSoCs. 
Building on RunSoC 1.0, which targeted allocation on homogeneous hardware, RunSoC 2.0 extends the framework to heterogeneous platforms by modeling processor-specific execution times, cluster-level organization, and domain-specific processing properties. It represents task sets as directed acyclic graphs (DAGs) subjected to strict end-to-end latency and core-affinity constraints, and formulates task scheduling and allocation as a multi-objective optimization problem that minimizes hierarchical memory-budget violations and inter-core/inter-cluster communication penalties. The framework supports multiple solving backends, including COIN-OR Branch and Cut (CBC), Google OR-Tools CP-SAT, and a Genetic Algorithm (GA), enabling comparative evaluation of exact, constraint-programming, and meta-heuristic approaches.
We evaluate RunSoC 2.0 using synthetic automotive task sets ranging from 10 to 500 tasks, mapped to representative heterogeneous MPSoCs, including the Renesas R-Car V4H, NVIDIA Jetson AGX Orin, and TI TDA4VM. The results show that RunSoC 2.0 can generate feasible and optimal schedules, expose architectural bottlenecks, and support rapid comparison of platform alternatives. Notably, CP-SAT consistently outperforms both CBC and the GA across tightly constrained hard real-time scheduling instances.
By incorporating cluster-aware communication and memory modeling, RunSoC 2.0 improves the realism of early-stage MPSoC analysis while retaining practical solution times for large automotive workloads. The framework therefore provides a practical basis for evaluating centralized automotive architectures under strict latency and resource constraints.


\end{abstract}
\section{Introduction}\label{sec:introduction}
Automotive E/E architectures are shifting from distributed networks of single-function Electronic Control Units toward centralized high-performance platforms. This transition is driven by the computational demands of advanced driver-assistance systems, autonomous driving, and infotainment workloads, which increasingly require heterogeneous Multi-Processor System-on-Chip (MPSoC) combining CPU, GPU, digital signal processors, and specialized accelerators \cite{configuringadas,deployingvisionbased,spatiotemporal}.
While consolidation can reduce wiring complexity and improve resource sharing, it also creates mixed-criticality environments in which safety-critical and best-effort tasks must coexist under strict timing, memory, and communication constraints \cite{configuringadas,contextinformation}.

These constraints make early-stage task allocation and scheduling a central design problem. Automotive workloads are commonly represented as precedence-constrained task graphs or Directed Acyclic Graphs (DAGs), whose tasks must be mapped to eligible processing elements while satisfying end-to-end latency bounds and avoiding resource contention \cite{energyaware,optimalPTGs}. On heterogeneous MPSoCs, this combined mapping and scheduling problem is computationally intractable in general, particularly when processor-specific execution times, affinity restrictions, and inter-core communication costs are considered \cite{configuringadas,dynamicpowermapping}.

RunSoC 1.0 \cite{mauser2026runsoc} addressed homogeneous task allocation heuristically while RunSoC 2.0 replaces this with multi-objective solver-based scheduling and allocation for heterogeneous MPSoCs. It models processor-specific execution times, cluster-aware memory and communication costs, task affinities, and end-to-end latency constraints, and compares COIN-OR Branch and Cut (CBC), OR-Tools CP-SAT, and a Genetic Algorithm (GA).

\section{Research Method and Related Work}\label{sec:research}
To design the architectural modeling and solver selection of RunSoC 2.0, we employed a mixed-methods approach comprising a rapid literature review \cite{cartaxo2020rapid} and semi-structured expert interviews \cite{hove2005experiences} conducted at Daimler Truck AG. The rapid review was structured using PICOC criteria and research questions to gain insights in task-core affinities, MPSoC partition properties, and scheduling/allocation approaches.
Synthesizing the literature and expert insights revealed two foundational requirements for our framework: first, metaheuristic approaches, such as GAs, show strong potential for overcoming the scalability bottlenecks of traditional exact scheduling; second, hierarchical memory contention is a dominant bottleneck that must be accounted for at design time.

The complete review protocol, including the PICOC criteria, search string evolution, inclusion/exclusion criteria, data extraction criteria, synthesized data, and derived feature implementation list, together with the interview guidelines, is available in our supplementary repository on \href{https://github.com/FromSWCtoSemi/RunSoC/tree/main/00_Rapid-Review/RunSoc_02\%20-\%20Heterogeneous}{GitHub} and \href{https://zenodo.org/records/20204691}{Zenodo}.

While early-design environments such as AMALTHEA APP4MC provide extensive modeling support for multicore systems, including software, hardware, timing, mapping, memory, and scheduling-related constraints,\footnote{\scriptsize\url{https://eclipse.dev/app4mc/}} they primarily serve as modeling and analysis infrastructures. In contrast, RunSoC 2.0 explicitly combines task allocation, scheduling, memory-budget violations, communication costs, and end-to-end latency constraints in a single solver-based optimization.

\section{RunSoC 2.0 Framework Overview}\label{sec:R2Overview}
RunSoC 2.0 follows four steps: task-set and platform description, problem parsing, feasibility pre-checking, and solver execution.
The exact pseudocode and formal models are available online on \href{https://github.com/FromSWCtoSemi/RunSoC/tree/main/03_Formal-Specifications/RunSoc_02\%20-\%20Heterogeneous}{GitHub} and \href{https://zenodo.org/records/20204691}{Zenodo}.

\subsection{Task Set and Platform Description}\label{sub:taskdesc}
The user describes the MPSoC platform and task set in a structured JSON input file. The platform model captures clusters, cores, execution-time scalars, memory budgets, supported execution domains, and optional communication paths. Heterogeneity of the platform stems from the potentially differing WCET scale per core, memory budgets and communication paths. Each task, which represents an indivisible schedulable unit of automotive software functionality, consists of a base execution time, memory demand, activation type (\textit{periodic}/\textit{event}), required execution domain or eligible cores, and predecessor relations.

RunSoC 2.0 distinguishes periodically triggered tasks from event-triggered tasks. Periodic tasks define task-chain releases, while event-triggered successors are activated by predecessor completion. Dependency-connected tasks form chains that must complete within the corresponding period, which is treated as an end-to-end latency constraint. Tasks may also be assigned execution-domain labels such as \textit{general-purpose}, \textit{safety}, \textit{sensor}, \textit{automotive-comms}, \textit{vision/accelerated}, or \textit{security}, provided that these labels match the domains offered by the platform cores.

Optional configuration parameters allow users to tune the optimization, including memory-penalty scaling, default communication penalties for unspecified communication paths, and maximum jitter bounds for task-chain releases.

\subsection{Problem Parsing and Feasibility Check}\label{sub:propfeas}
The framework accepts JSON input through a Python Flask solver endpoint, validates it, and maps it to an internal problem instance. Task chains are inferred from periods and dependency relations. The scheduling horizon is derived as the least common multiple of all chain period.

Since tasks within a periodic task chains are executed repeatedly, RunSoC 2.0 generates jobs which represent a schedulable instance of a task. For each task with period $p_t$, the number of jobs over horizon $H$ is $\max\left(0,\left\lceil H/p_t \right\rceil\right)$. Before invoking the main solver, a feasibility pre-check abstracts the instance into a task-to-core bin-packing problem with per-core utilization bounds. The resulting balanced assignment minimizes maximum core utilization and can be passed to the main solver as an initial hint.

\subsection{Scheduling and Allocation}\label{sub:schedalloc}
The scheduling and allocation logic, including all hard and soft constraints, is preserved across all three solvers. By employing a partitioned scheduling approach, allocation is resolved strictly at a task level. Consequently, all constituent jobs inherit their parent task's core assignment. The objective function minimizes a weighted sum of \textbf{(1) Memory Overflows:} Exceedances of soft memory budgets at both the core and cluster levels and \textbf{(2) Communication Penalties:} Cost incurred by data dependencies between tasks, scaled by hardware locality (intra-core, inter-core, or inter-cluster).

The optimization is subject to hard constraints for exact task assignment, task-level resource accounting, release and deadline bounds, precedence preservation, and non-overlap of jobs assigned to the same core. Memory demand is counted once per task assignment, independent of the number of generated jobs, while timing constraints apply to each concrete job instance.

While the Integer Linear Programming (ILP) formulation could be translated to CBC and CP-SAT libraries with relative ease, GA is a metaheuristic that requires solutions to be encoded as chromosomes.

\section{Experimental Design}
Since production automotive task sets are largely proprietary \cite{waters}, we evaluate RunSoC 2.0 using synthetic DAG workloads derived from the statistical properties of the WATERS15 benchmark \cite{waters}.

\subsection{Workload Generation}
The synthetic workloads are designed to accurately reflect the structure, data dependencies, and timing constraints of real-world automotive control. To thoroughly test the scalability of our framework, we generated task sets of varying complexities, comprising 10, 25, 50, 100, 200, and 500 tasks. 

Following the WATERS15 benchmark characteristics, the worst-case execution times (WCET) of the generated tasks are modeled using a Weibull distribution, parameterized according to realistic minimum, average, and maximum execution times \cite{waters}. To test the framework's ability to handle hard real-time requirements, the generated DAGs feature specific cause-effect chains.

\subsection{Target Hardware Platforms}
The synthetic task sets are mapped and scheduled onto hardware models of three representative, state-of-the-art heterogeneous MPSoCs: the Renesas R-Car V4H, NVIDIA Jetson AGX Orin, and Texas Instruments TDA4VM. The architectural parameters for these platforms were extracted directly from their publicly available data sheets.

By modeling these specific platforms, RunSoC 2.0 incorporates processor-specific domain properties (the ones described in \ref{sub:taskdesc}) and cluster-level organization. This allows the experimental setup to enforce realistic core-affinity constraints and evaluate the framework's multi-objective optimization, which specifically targets the minimization of hierarchical memory-budget violations and the communication penalties incurred when data is transferred across different cores or clusters.

\subsection{Solver Configuration and Execution}
Exact optimization approaches such as ILP can suffer from exponential runtime growth as task and processor counts increase. Therefore, each solver run was limited to 1000 seconds. Runs that did not return a feasible solution within this timeout were counted as unsuccessful in the feasibility analysis and omitted from runtime plots requiring completed results. Thus, missing data points indicate timeout-related non-completion rather than proven infeasibility.

The experiments compare each solver's ability to return feasible, optimal, or near-optimal schedules within the timeout, thereby evaluating CP-SAT and GA against the ILP-based CBC formulation.

\section{Results}
The complete dataset, including all synthetic task sets, platform configurations, raw solver outputs, and supplementary analysis scripts, can be found on \href{https://github.com/FromSWCtoSemi/RunSoC/blob/main/04_Evalution/data.zip}{Github} and \href{https://zenodo.org/records/20204691}{Zenodo}.

\begin{figure}[htbp]
  \centering

  \begin{subfigure}[t]{0.495\textwidth}
    \centering
    \includegraphics[
      width=\linewidth,
      height=0.38\textheight,
      keepaspectratio
    ]{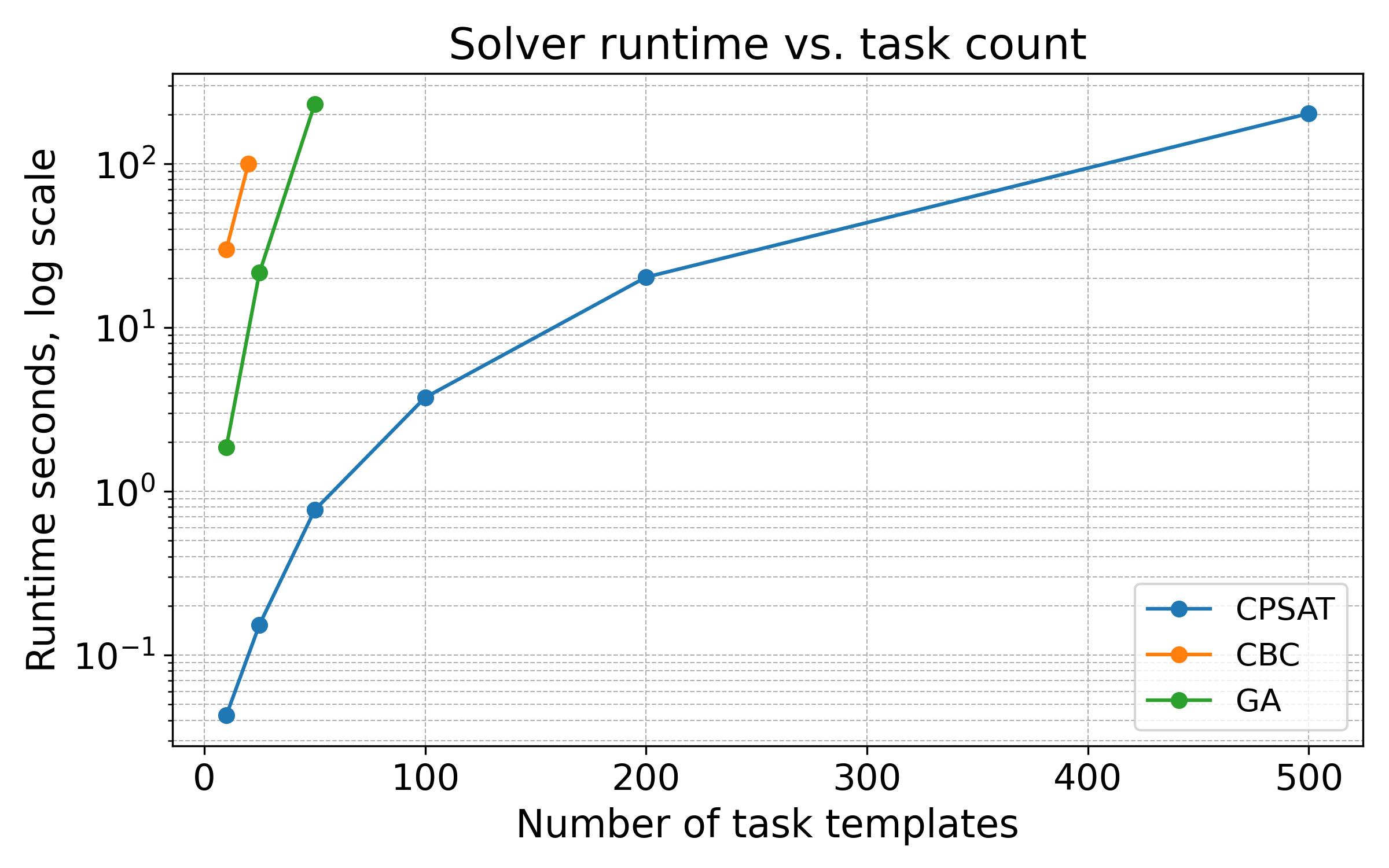}
    \caption{Runtime by task count.}
    \label{fig:runtime-task-count}
  \end{subfigure}
  \hfill
  \begin{subfigure}[t]{0.495\textwidth}
    \centering
    \includegraphics[
      width=\linewidth,
      height=0.38\textheight,
      keepaspectratio
    ]{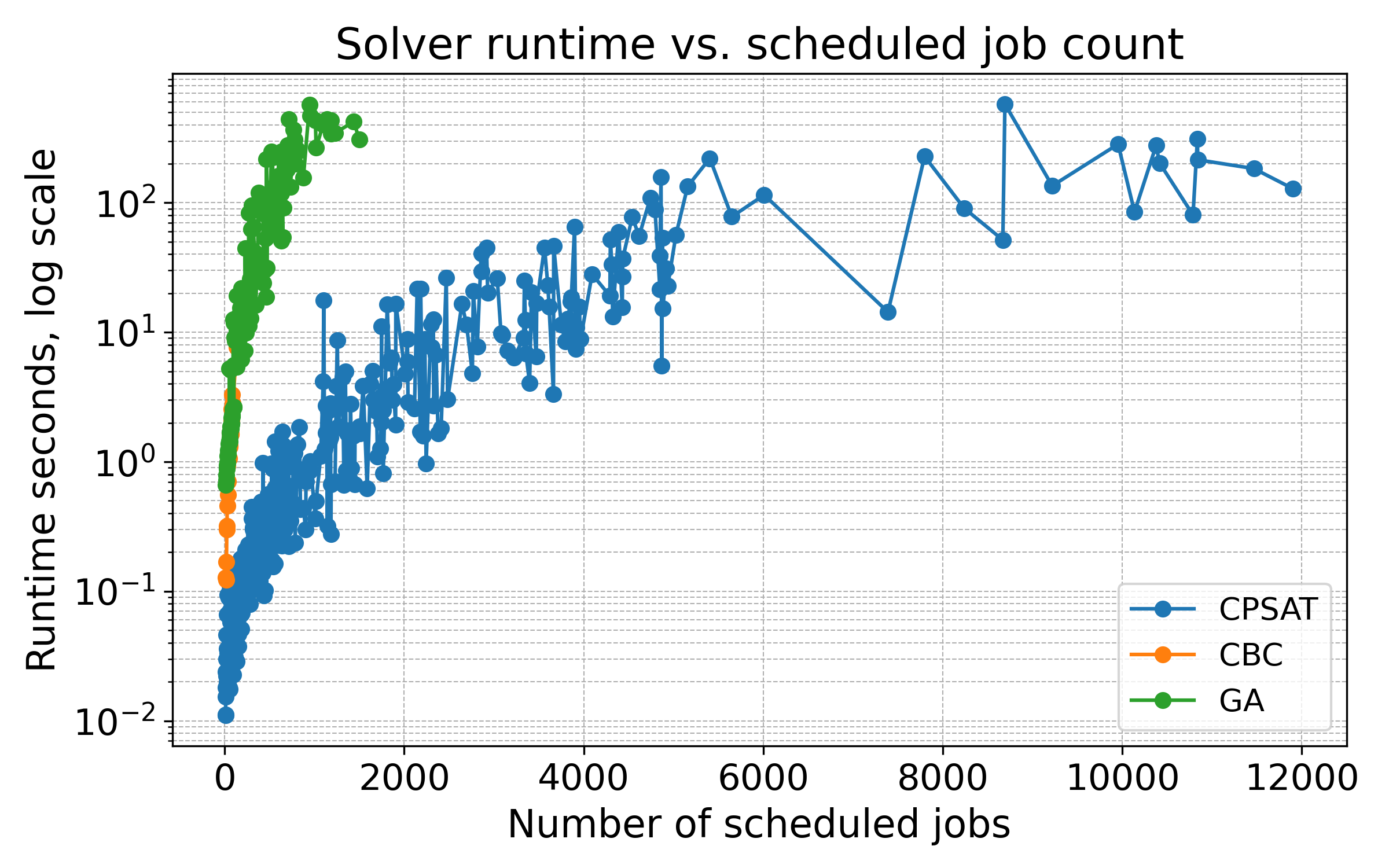}
    \caption{Runtime by expanded job count.}
    \label{fig:runtime-job-count}
  \end{subfigure}

  \caption{Runtime scalability under the imposed timeout. CP-SAT remains operational over larger task and job counts than CBC and GA. Missing points indicate solver runs that did not return a completed result before the timeout.}
  \vspace{-0.5em}
  \label{fig:runtime-scalability}
  \vspace{-0.5em}
\end{figure}

\begin{figure}[htbp]
  \centering
  \includegraphics[
    width=0.75\textwidth,
    height=0.40\textheight,
    keepaspectratio
  ]{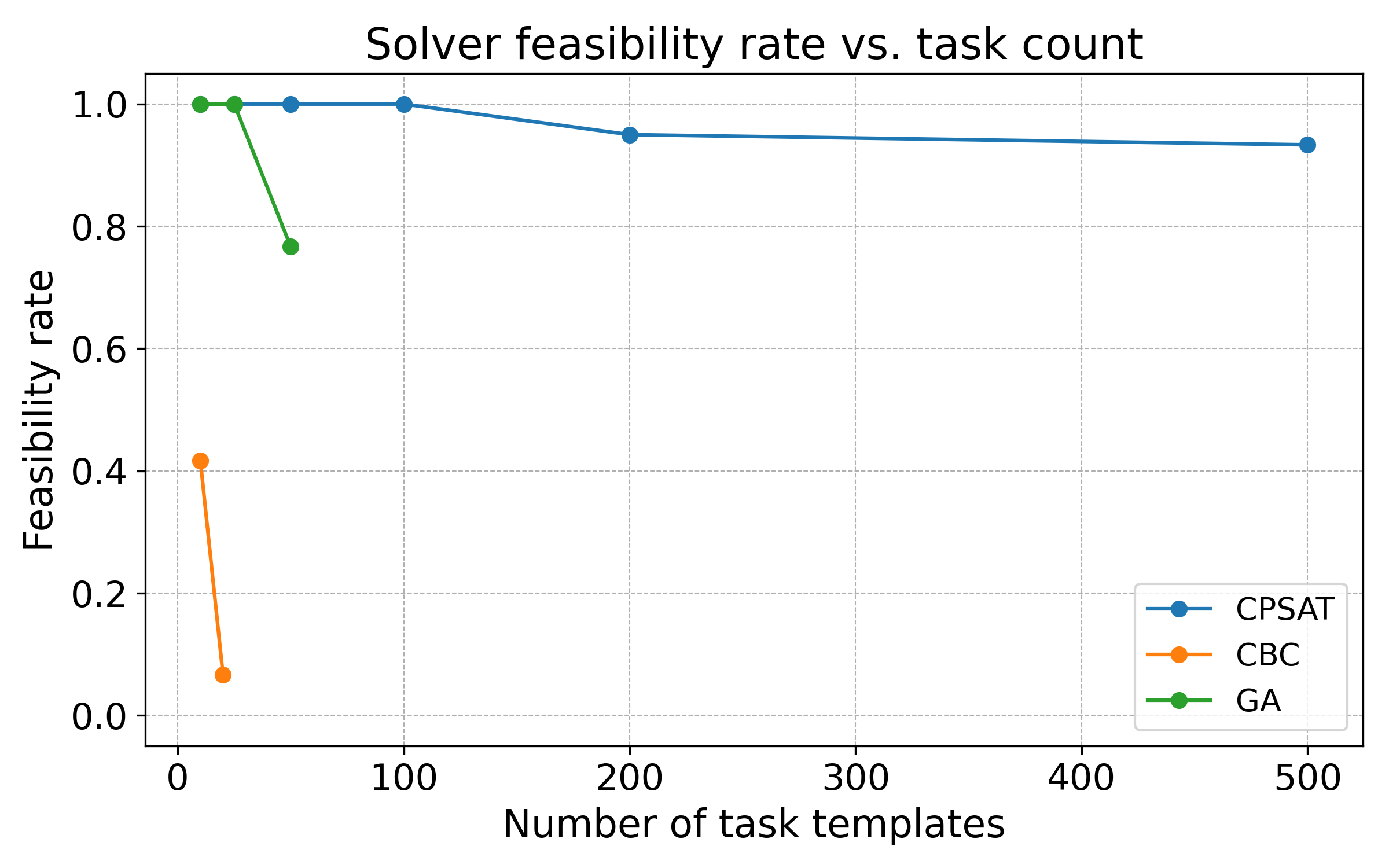}
  \vspace{-0.5em}
  \caption{Feasibility rate over attempted instances. CP-SAT retains high feasibility up to 500 tasks; CBC and GA return feasible schedules only for smaller instances. Missing points indicate solver runs that did not return a feasible schedule before the timeout, not necessarily proven infeasibility.}
  \label{fig:feasibility-rate}
  \vspace{-0.5em}
\end{figure}

Figures~\ref{fig:runtime-task-count} and~\ref{fig:runtime-job-count} report runtime scalability, while Figure~\ref{fig:feasibility-rate} shows the fraction of instances for which each solver returned a feasible schedule. These metrics should be interpreted jointly: for CBC and GA, low coverage at larger task sizes mainly indicates timeout-related failure to return feasible schedules, not evaluation on the same instance population as CP-SAT.

Overall, CP-SAT was the only solver that scaled to the full benchmark range. It solved all instances up to 100 tasks, retained 95.0\% feasibility at 200, and still returned feasible schedules for 93.3\% of the 500-task cases. Its median runtime rose from 0.043s at 10 tasks to 202.71s at 500, remaining under the 1000s timeout.

GA and CBC showed substantially weaker scalability. GA was evaluated successfully only up to 50 tasks and already required median runtimes of 1.86s, 21.58s, and 231.44s for 10, 25, and 50 tasks, respectively. Thus, although GA can return feasible schedules on small instances, its runtime growth prevents it from serving as a competitive backend for larger scheduling problems under the current configuration. CBC performed worst in this setting. It reached the imposed time limit on most runs and was only usable on very small instances, with median runtimes of 30s and 100s for 10 and 25 tasks, respectively. Consequently, CBC is not competitive for the evaluated formulation and timeout regime.

The job-count view in Fig.~\ref{fig:runtime-job-count} confirms that the scalability advantage of CP-SAT persists after expanding periodic chains into concrete jobs, while GA and CBC terminate at substantially smaller schedules.

Objective-value comparisons must be interpreted cautiously because CBC and GA do not cover the same instance set as CP-SAT. Since both solvers fail to return feasible schedules for many medium and large instances, aggregate objective comparisons across all task sizes would be right-censored and affected by survivorship bias. In other words, CBC and GA objective values are available mostly for smaller or easier instances, while CP-SAT also covers larger and harder instances. Therefore, feasibility and runtime are the primary scalability metrics in this evaluation. Objective quality is considered only for instances where a solver returned a feasible schedule.

On commonly solved small instances, CP-SAT consistently matched the best-known objective values obtained in the experiment. CBC occasionally matched these values as well, but only on the small subset of instances for which it returned a feasible solution. GA produced feasible schedules on small and medium-small instances, but its solution quality degraded as task count increased. This indicates that GA's main limitation is twofold: it both scales poorly in runtime and increasingly fails to converge to competitive schedules under the fixed timeout.

The diagnostic objective components show that memory pressure dominates the soft-constraint costs, while communication penalties are negligible in most returned schedules. For CP-SAT, the median memory penalty increased from 82 at 10 tasks to 85,820 at 500. This suggests that the evaluated workloads and platform models are mainly memory-constrained. However, this is a property of the current benchmark and penalty calibration, not a general claim about all heterogeneous MPSoC deployments.

In summary, CP-SAT provides the best feasibility, runtime, and solution-quality trade-off, scaling to 500 tasks and several thousand jobs. GA solves only small instances competitively, while CBC frequently times out even on small cases. Overall, memory pressure dominates the objective, motivating future work on memory-aware mapping, platform memory modeling, and penalty calibration.

\section{Conclusion and Outlook}\label{sec:CO}
This paper presented RunSoC 2.0, a framework for early-stage scheduling and allocation of automotive software tasks on heterogeneous MPSoCs. The framework advances beyond RunSoC 1.0 \cite{mauser2026runsoc} by replacing its heuristic homogeneous-allocation model with multi-objective scheduling and allocation over processor-specific execution times, task-to-core affinities, cluster-aware communication costs, hierarchical memory budgets, and end-to-end latency constraints. The evaluation on synthetic automotive workloads and representative MPSoCs shows that RunSoC 2.0 can generate feasible schedules, compare solver backends, and expose architectural bottlenecks early in the design process.

Among the evaluated solvers, CP-SAT showed the strongest overall behavior, scaling to the largest task sets while maintaining practical runtimes and high solution quality. CBC’s low returned-feasible-solution rate mainly reflects early timeout behavior rather than fundamentally different feasibility, while the GA was limited by degraded runtime and solution quality on larger instances. Across platforms, memory pressure emerged as the dominant bottleneck, indicating that memory-aware mapping is essential for centralized automotive MPSoCs.

Future work will improve the fidelity of the platform model by incorporating memory-bandwidth constraints, bus contention, and finer-grained communication semantics such as shared-memory and DMA-assisted transfers. We also plan to model sporadic and interrupt-driven tasks, mixed-criticality constraints, and cache-related preemption delays. Finally, future evaluations should include industrial task sets and hybrid solving strategies that combine CP-SAT with heuristic warm starts or decomposition techniques.

\subsubsection*{Declaration on the Use of Generative AI}
Generative AI tools were used for text review and formatting. The authors remain fully responsible for the manuscript and code artifacts.

\bibliographystyle{splncs04}
\bibliography{mybibliography}

\end{document}